\documentclass[preprint]{elsarticle}

\usepackage{amssymb}
\usepackage{amsmath}
\usepackage{graphicx}

\def\p{{\boldsymbol p}}

\def\qqb{{q\bar q}}
\def\pp{{\boldsymbol p}}
\def\ppb{{\bar {\boldsymbol p}}}
\def\q{{\boldsymbol q}}
\def\n{{\boldsymbol n}}
\def\a{{\boldsymbol a}}

\def\k{{\boldsymbol k}}

\def\r{{\boldsymbol r}}

\def\IinterfOne{\mathcal{I}^\text{interf\,I}_\qqb}
\def\IinterfTwo{\mathcal{I}^\text{interf\,II}_\qqb}
\def\Iindep{\mathcal{I}^\text{indep}_q}
\def\Icoh{\mathcal{I}^\text{coh}_q}

\newcommand{\beq}{\begin{eqnarray}}
\newcommand{\eeq}{\end{eqnarray}}

\long\def\comment#1{ }    %% Text in {} not included in dvi file.
\newcommand{\be}{\begin{equation}}
\newcommand{\ee}{\end{equation}}

\begin{document}

\begin{frontmatter}

\title{Coherence effects and broadening in medium-induced QCD radiation off a massive $q\bar q$ antenna}

\author[usc]{N. Armesto}

\author[usc]{H. Ma}

\author[usc]{Y. Mehtar-Tani}

\author[usc]{C. A. Salgado}

\author[lu]{and K. Tywoniuk}

\address[usc]{Departamento de F\'isica de Part\'iculas and IGFAE, 
Universidade de Santiago de Compostela, 
E-15782 Santiago de Compostela, 
Galicia-Spain}

\address[lu]{Department of Astronomy and Theoretical Physics, 
Lund University, S\"olvegatan 14A,
S-223 62 Lund, 
Sweden}

\begin{abstract}

Studies of medium-induced QCD radiation usually rely on the calculation of single-gluon radiation spectrum off an energetic parton traversing an extended colored medium. Recently, the importance of interference effects between emitters in the medium has been explored. In this work we extend previous studies by calculating the single-gluon coherent spectrum off an antenna consisting of a massive quark-antiquark pair. Interferences dominate the spectrum of soft gluons, which are mainly emitted outside of the cone made by the antenna opening angle, while the antenna  results in a superposition of independent emitters above a critical gluon energy scale. We study the interplay between the dead-cone effect and medium-induced jet broadening in both cases of soft and hard gluons and present results on energy loss distributions. 

\end{abstract}

\begin{keyword}
Perturbative QCD, Medium-induced radiation, Jet quenching
\end{keyword}

\end{frontmatter}

\section{Introduction}

A wide range of observations suggest that a hot and dense state, wherein quarks and gluons are the relevant degrees of freedom, is formed in ultrarelativistic heavy-ion collisions. In particular, processes involving a large momentum scale, so-called hard probes, are useful in pinpointing the characteristics of this new state of matter, eventually a quark-gluon plasma (QGP). Since they occur very early on in the collisions, their production cross section is taken to be the same as in vacuum while their subsequent fragmentation is assumedly sensitive to such properties as the temperature and local density of the evolving dense medium. 

One of the observables belonging to this class is simply the yield of light particles with large transverse momenta produced in heavy-ion collisions. Experiments at RHIC \cite{Adcox:2004mh, Back:2004je, Arsene:2004fa, Adams:2005dq, Adare:2008qa} and LHC \cite{Aamodt:2010jd} have measured this yield to be almost a factor 5-6 times smaller than expected had the nuclear collision simply been a superposition of independent nucleon-nucleon collisions. Besides, back-to-back correlations are distorted and washed out \cite{Adams:2006yt}. This striking phenomenon has been dubbed ``jet quenching". Furthermore, the suppression of decay products originating from heavy quarks, such as non-photonic electrons  \cite{Abelev:2006db, Adare:2006nq} or reconstructed charmed mesons \cite{Dainese:2011vb},  seem to follow the same pattern and is comparable in magnitude. This suggests a significant medium-modification of the fragmentation pattern of highly energetic partons that traverse the dense medium. Recently, studies of fully reconstructed jets produced in heavy-ion collisions have been carried out at both RHIC \cite{Putschke:2008wn, Salur:2008hs, Bruna:2009em, Ploskon:2009zd} and LHC \cite{Aad:2010bu, Chatrchyan:2011sx}. In this context, tagging heavy-flavored jets is a tool to study the mass and color dependence of partonic energy loss \cite{Lokhtin:2004rg, Armesto:2005iq}. 

To date, the standard theoretical explanation of these phenomena is radiative energy loss due to medium-induced gluon radiation, for details see the recent reviews \cite{CasalderreySolana:2007zz,d'Enterria:2009am, Wiedemann:2009sh, Majumder:2010qh}. Studies have shown that a mass ordered suppression pattern, where the light partons are most strongly affected, is expected from medium-induced radiative processes \cite{Armesto:2005iq}. But the similar suppression observed for light hadrons and non-photonic electrons at RHIC presents problems for its description within radiative energy loss models and other mechanisms have been invoked, see \cite{Armesto:2009zi} and refs. therein. Studies of correlations with tagged heavy-quark jets might help in disentangling the possible effects and their interplay. The mass dependence of the suppression serves in any case as a useful discriminator between energy loss approaches.

From perturbative QCD one generically expects the radiative process off a heavy parton to be less intensive than off a massless one. In vacuum, this is simply due to the non-zero mass of the emitter which cuts off the collinear singularity at an angle $\theta_0 = m_q/E_q$, dubbed the dead-cone angle \cite{Dokshitzer:1991fc}, where $m_q$ and $E_q$ are the mass and the energy of the radiating parton, respectively. Besides, a non-zero mass reduces the formation time of the emitted gluon, given by $t_q \sim \big(\omega(\theta^2 + \theta_0^2)\big)^{-1}$, where $\omega$ and $\theta$ are the energy and angle of the emitted gluon. The dead-cone suppression also plays an important role in the presence of a medium \cite{Basics of Perturbative QCD, Armesto:2003jh, Djordjevic:2003zk, Zhang:2003wk}, especially for energetic gluons which populate small emission angles. The latter effect, on the other hand, implies a relative enhancement of the medium-induced radiation as the intensity is roughly proportional to the inverse of the formation time. In other words, this relative enhancement occurs as a result of the destructive interference between coherent scattering centers in the medium, acting as a single scattering within the gluon formation time, being repressed. In general, the net result of this competition is that massive partons lose less energy in a medium than massless ones.

Considering induced gluon radiation as the only mechanism for in-medium energy loss, the fact that a heavy quark loses as much energy as a light one indirectly implies that the radiation that takes away the energy necessarily has to occur at large angles since the dead-cone suppression only plays a role for small-angle radiation. In fact, large-angle radiation is a hallmark feature of the medium-induced spectrum which follows inherently from the gluonic nature of the interaction with the colored medium. Recently, an additional contribution to large-angle radiation, particularly important for soft gluons, was found \cite{MehtarTani:2010ma, MehtarTani:2011tz, MehtarTani:2011jw, CasalderreySolana:2011rz}: whereas the medium-induced single-gluon spectrum is both collinear and infrared finite, interferences between fragments of a jet, e.g., constituting a quark-antiquark antenna, result in a soft divergence which induces a logarithmically enhanced radiative component which is delimited to angles larger than the opening angle of the pair. This mechanism also controls the breakdown of angular ordering for soft gluons in the medium. 

In this work we generalize previous results to the case of radiation off a massive antenna traversing a static-colored medium and study the energy loss distributions as a function of the relevant kinematic variables. For the sake of transparency, presently we only allow for a single scattering with the medium. We find that the radiation spectra off the antenna constituents are mainly affected by the dead-cone suppression in the presence of a dense medium. For smaller parameters and less favourable geometrical situations, however, the mass ordering can be reversed. These trend parallel results from earlier studies of the behavior of the independent spectra off a single constituent \cite{Armesto:2003jh}. This is due to the fact that the energy loss distributions are biased, in this parametric situations, toward hard gluon emissions, a regime where all interferences are of minor importance. 

Restricting our study to the soft sector, i.e. calculating how much energy is taken by soft gluons, we find an enhancement for the antena spectrum compared to the independent one, in line with previous findings. The dead-cone suppression decreases with the opening angle of the pair up to a characteristic cut-off, where interferences between emitters start to drop. This situation results from the fact that vacuum radiation is heavily affected inside the cone, while the medium-induced radiation is insensitive because it is restricted to large angles. In effect, the dead-cone suppression is removed for soft gluons at large opening angles of the pair.

The paper is organized as follows. In Section 2, the explicit expression of the medium-induced gluon radiation spectrum off a massive antenna is given and its soft limit is discussed. In Section 3, numerical results for the gluon spectrum,  average energy loss and broadening are given and are discussed, together with the transition between the interference spectrum and the superposition of independent spectra. Conclusions are provided in Section 4. Some technical details of the derivations are given in the Appendices.

\section{Medium-induced gluon radiation spectrum}

\begin{figure}[t]
\begin{center}
\includegraphics[width=0.8\textwidth]{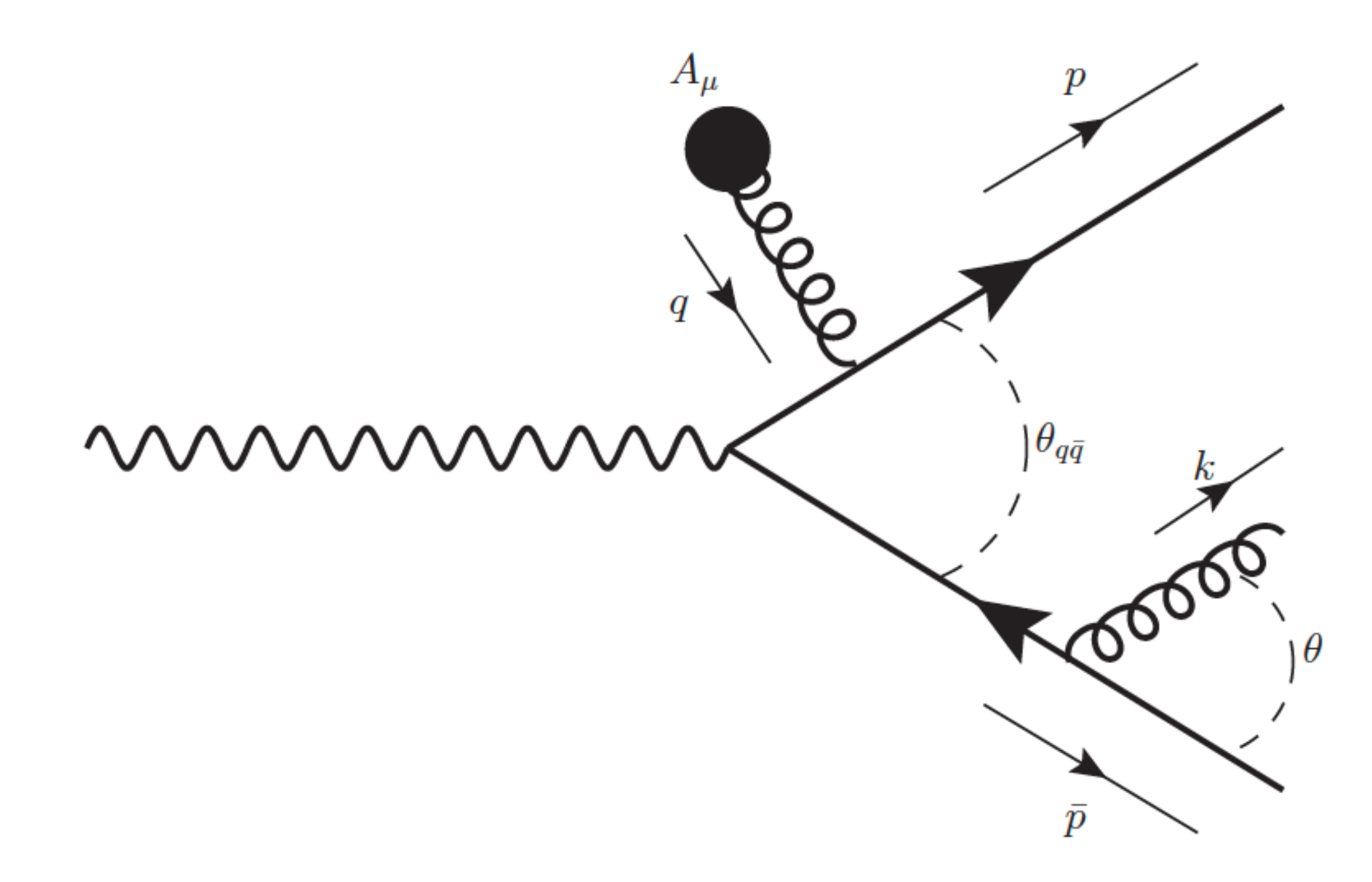}
\caption{\label{fig:kin} An example diagram demonstrating the notation for the momenta.}
\end{center}
\end{figure}

We study the spectrum off an antenna made out of a massive quark-antiquark ($q\bar q$) pair traversing a colored medium assumed to be made out of static scattering centers spaced, on average, at a typical mean free path $\lambda$. The medium interaction is modeled by a classical vector field $A_\mu(q)$, where $q$ denotes the momentum exchange. The single elastic differential scattering cross section $\propto \left| A(q)\right|^2$ is then usually chosen to be a Yukawa potential screened by the Debye mass $m_D$. For the sake of transparency, we study the case when the antenna arises from the decay of a highly virtual time-like photon\footnote{The calculation of the decay of a virtual gluon proceeds, in the soft limit, analogously to the one performed in this work, differing only in the color algebra.}. The quark and antiquark kinematics is given by $p\equiv(E_q,\vec p)$ and $\bar p \equiv (E_{\bar q}, \vec{\bar p} )$, respectively, where we put $E_q=E_{\bar q}$. The 3-momentum is given by $\vec p \equiv (\pp,p_z)$ and $\vec {\bar p} \equiv (\ppb, \bar p_z)$, as usual, and $p^2 = \bar p^2 = m_q^2$ is the mass of the quarks. The $\qqb$ antenna is characterized by the opening angle $\theta_\qqb$. Finally, either of the antenna constituents can radiate a gluon with 4-momentum $k\equiv(\omega,{\bf k})$, the gluon emission angle being $\theta$. The physical process under consideration is illustrated in Fig.~\ref{fig:kin}.

\begin{figure}[t]
\begin{center}
\includegraphics[width=\textwidth]{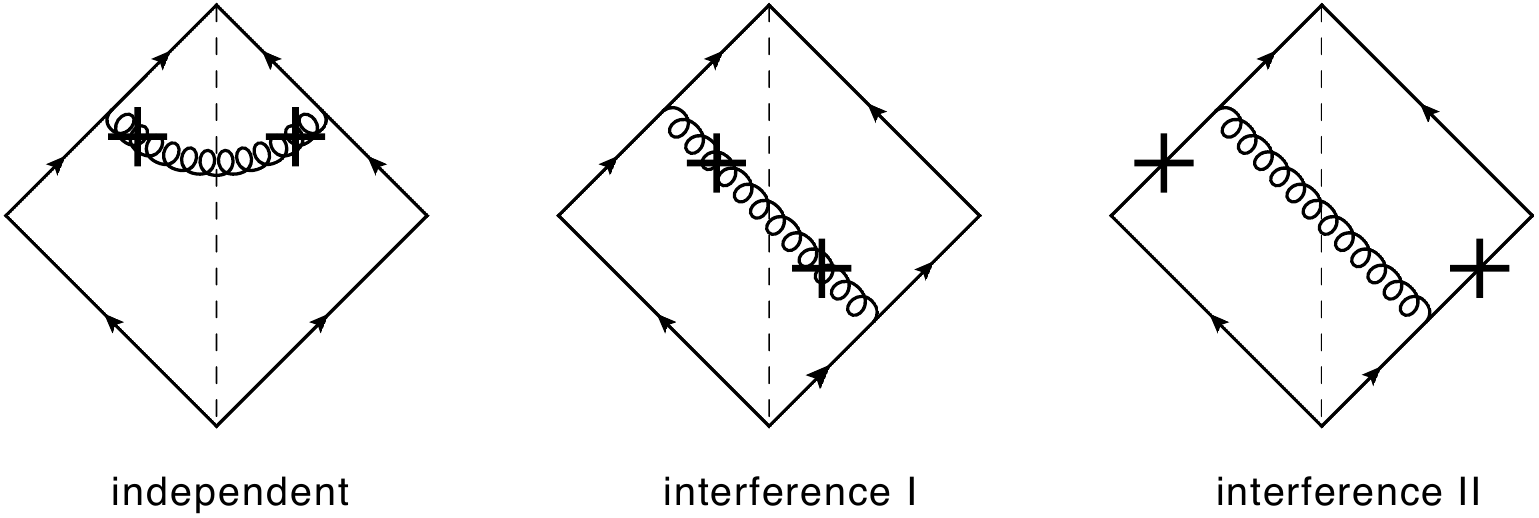}
\caption{\label{fig:aspec} An illustration of the three contributions to the antenna spectrum in Eq.~(\ref{eq:master2}), characterizing typical situations: independent emissions with gluon rescattering and interference contributions with and without gluon rescattering. 
}
\end{center}
\end{figure}

Since the Born cross section is not altered by the presence of the medium, it factors out of the expression. Then, the single-inclusive gluon spectrum off the antenna can be decomposed in terms of three characteristic contributions, which we denote
\beq
(2 \, \pi)^3 2 \, k^+ \frac{{\rm d}N}{{\rm d}^3 k_{\rm LC}} = {\cal I}_{q {\bar q}}^{\rm indep} + {\cal I}_{q {\bar q}}^{\rm interf \, I} + {\cal I}_{q {\bar q}}^{\rm interf \, II} \,,
\label{eq:master2}
\eeq
where $k^+$ is the gluon light-cone energy and ${\rm d}^3 k_{\rm LC}$ $=$ ${\rm d} k^+ {\rm d}^2 {\bf k}$ is the light-cone variable of integration. For technical details of the calculation, we refer the reader to \ref{sec:appa} and \cite{Wiedemann:2000ez, Gyulassy:2000er, CasalderreySolana:2007zz}. The three contributions in Eq.~(\ref{eq:master2}) are illustrated by three example diagrams in Fig.~\ref{fig:aspec}. They represent the independent emission off the quark and the antiquark, contained in ${\cal I}_{q {\bar q}}^{\rm indep}$, and the interferences, given by ${\cal I}_{q {\bar q}}^{\rm interf \, I} + {\cal I}_{q {\bar q}}^{\rm interf \, II}$, i.e., contributions where the gluon is emitted by one of the antenna constituents to the left of the cut and absorbed by the other to the right. We go on to discuss the characteristic features of these three contributions in the following subsections.

\subsection{The independent emission spectrum}
\label{sec:independent}

Let us turn to the sum of independent emission spectra off the quark and the antiquark, denoted as  ${\cal I}_{q {\bar q}}^{\rm indep}$ above\footnote{This part of the spectrum was denoted as ``GLV" in previous works of some of the authors \cite{MehtarTani:2010ma}.}. This contribution reads
\begin{equation} 
\label{eq:GLV}
\begin{split}
{\cal I}_{q {\bar q}}^{\rm indep} = & \, 2 \,(4\pi)^2 \alpha_s^2  \, C_A \, C_F\, n_0\, \int_0^{L^+} {\rm d} x^+  \, \int \frac{{\rm d}^2\q}{(2 \, \pi)^2} \, \frac{m _D^2}{(\q^2 + m _D^2)^2}\\
& \bigg[\bigg(\frac{{\boldsymbol \nu}^2}{(x\, p \cdot v)^2} - \frac{{\boldsymbol \nu} \cdot {\boldsymbol \kappa}}{x^2(p \cdot v) \, (p \cdot k)}\bigg) \Big(1 - \cos\Omega_q \, x^+\Big) \\
& + \bigg(\frac{{\bar {\boldsymbol \nu}}^2}{(\bar x \, {\bar p} \cdot v)^2} - \frac{{\bar {\boldsymbol \nu}} \cdot {\bar {\boldsymbol \kappa}}}{\bar x^2 ({\bar p} \cdot v) \, ({\bar p} \cdot k)}\bigg) \Big(1 - \cos\Omega_{\bar q} \, x^+\Big)\bigg] \,,
\end{split}
\end{equation}
where all momenta and positions are written in light-cone coordinates (LC), see  \ref{sec:appa} for details. Indeed, ${\cal I}_{q {\bar q}}^{\rm indep} $ is a superposition of the individual spectra off the quark and the antiquark, which are to be found in the second and third line of Eq.~(\ref{eq:GLV}), respectively, and therefore does not contain information about the opening angle of the pair. Above, we assumed a medium of constant density $n_0$ which extends over a distance $L^+=\sqrt{2} \, L$ in the longitudinal direction, such that $n_0 \, L^+ = L/\lambda$ gives the average number of scattering centers. 

In Eq.~(\ref{eq:GLV}), $\alpha_s C_F$ denotes the emission strength in terms of the strong coupling constant, and analogously $\alpha_s C_A$ for the interaction with the medium ($C_F=4 / 3$ is the Casimir factor of the fundamental representation of SU(3) while $C_A=3$ is the one in the adjoint representation). The momentum vector of the emitted interacting gluon is defined as $v \equiv [k^+, (\k - \q)^2 / (2 \, k^+), {\k} - {\q}]$ in LC. ${\boldsymbol \kappa}={\k} - x \, {\p}$ and ${\boldsymbol \nu}=({\k} - {\q}) - x \, \p$ are the transverse displacement vectors, where $x = k^+/p^+$, and $\Omega_q=p \cdot v / p^+$ denotes the inverse formation time. Analogous expressions hold for the antiquark contribution simply by substituting $p \to \bar p$. Kinematically, the terms proportional to ${\boldsymbol \nu}^2 / (x\, p \cdot v)^2$ account for the contribution purely from gluon rescattering, while those proportional to ${\boldsymbol \nu} \cdot {\boldsymbol \kappa} / \big(x^2(p \cdot v) \, (p \cdot k)\big)$ account for the contribution from the destructive interference between gluon and quark (antiquark) rescatterings. The contributions from purely quark rescattering cancel due to the contact terms, see \cite{Wiedemann:2000ez, Gyulassy:2000er}.

The independent spectrum off the quark simplifies in the $|{\p}|=0$ frame, where it reads
\begin{equation} \label{eq:GLVq}
\begin{split}
{\cal I}_{q}^{\rm indep} = &\, 8 \,(4\pi)^2 \alpha_s^2  \, C_A \, C_F\, n_0\, \int_0^{L^+} {\rm d} x^+  \, \int \frac{{\rm d}^2\q}{(2 \, \pi)^2} \, \frac{m _D^2}{(\q^2 + m _D^2)^2} \\
& \frac{({\k} - {\q})^2 \, {\k} \cdot {\q} - x^2 \, m_q^2 \, ({\k} - {\q}) \cdot {\q}}{\left[({\bf k} - {\q})^2 + x^2 \, m_q^2\right]^2 \, \left({\k}^2 + x^2 \, m_q^2 \right)} \bigg[1 - \cos\bigg(\frac{({\k} - {\q})^2 + x^2 \, m_q^2}{2 \, k^+} x^+\bigg)\bigg] \,.
\end{split}
\end{equation}
This expression coincides with the one obtained in \cite{Armesto:2003jh}. We recover the independent spectrum off a massless quark, which was first calculated in \cite{Wiedemann:2000ez, Gyulassy:2000er}, by setting $m_q=0$ in Eq.~(\ref{eq:GLVq}). 

Qualitatively, the interaction with the medium leads to a characteristic broadening of the mean transverse momentum, $\langle \k^2\rangle \sim m_D^2$, which further implies a characteristic energy scale. In the single scattering case, this characteristic gluon energy is denoted by $\bar \omega_c = m_D^2 L/2$ \cite{Armesto:2003jh,Salgado:2003gb}. The presence of these intrinsic scales in the medium-induced spectrum renders it both infrared and collinear safe, in contrast to the vacuum spectrum. 

As expected, the non-zero quark mass appears both in the effective formation time, in the argument of the cosine function, and as a dead-cone factor analogously to the vacuum case in Eq.~(\ref{eq:GLVq}). For soft gluons, these two modifications work in opposite directions and compensate each other. On the other hand, the dead-cone suppression at small angles is particularly important for the hard gluons, which mainly occupy this part of the phase space. Since the energy loss distribution is biased toward the hard sector, it follows that the typical energy loss is smaller in the massive than in the massless case \cite{Armesto:2003jh}. The non-zero quark mass does not change the infrared and collinear behaviors of the independent spectrum.

The individual spectra off the two antenna constituents lack information about the presence of the other participant. Assuming a strong medium screening, leaving the participants completely unaware of one another, it can serve as a building block for a cascade of multiple independent medium-induced gluon emissions \cite{qpythia1,qpythia2}. Yet, such a heuristic generalization fails to include subtle interference effects which are crucial for soft gluons, most prominently in vacuum \cite{Basics of Perturbative QCD}, and which were calculated recently \cite{MehtarTani:2010ma, MehtarTani:2011tz, MehtarTani:2011jw, CasalderreySolana:2011rz} for the medium. We now turn to the discussion of these in the following subsections.

\subsection{Medium-induced interferences}
\label{sec:interferences}
The main difference between ${\cal I}_{q {\bar q}}^{\rm indep} $ in Eq.~(\ref{eq:GLV}) and the complete antenna spectrum is the existence of the extra novel contributions stemming from the gluon exchange between the emitters of the antenna, i.e., the quark and the antiquark. 
These can be separated further according to their infrared behaviors, what we will call in the following type ``I" and type ``II" --- the former is infrared safe while the latter is infrared divergent. This has a simple diagrammatic interpretation, visualized in Fig.~\ref{fig:aspec} \, \footnote{This diagrammatic interpretation is in the spirit of classical currents, valid for soft emissions, and strictly speaking not in terms of Feynman diagrams due to the existence of the contact terms. See \ref{sec:appa} and \cite{Wiedemann:2000ez, Gyulassy:2000er, CasalderreySolana:2007zz} for further details.}. Namely, all interferences of type ``I" include at least one gluon interaction with the medium, while in case of type ``II" the gluon does not interact. The interaction of the off-shell gluon with the medium screens both the soft and collinear divergences in ${\cal I}_{q {\bar q}}^{\rm interf \, I}$, as is the case for the independent spectrum. In the soft limit, it is therefore the interference spectrum ``II" that gives the dominant contribution.
 
\begin{figure}
\begin{center}
\includegraphics[width=0.9\textwidth]{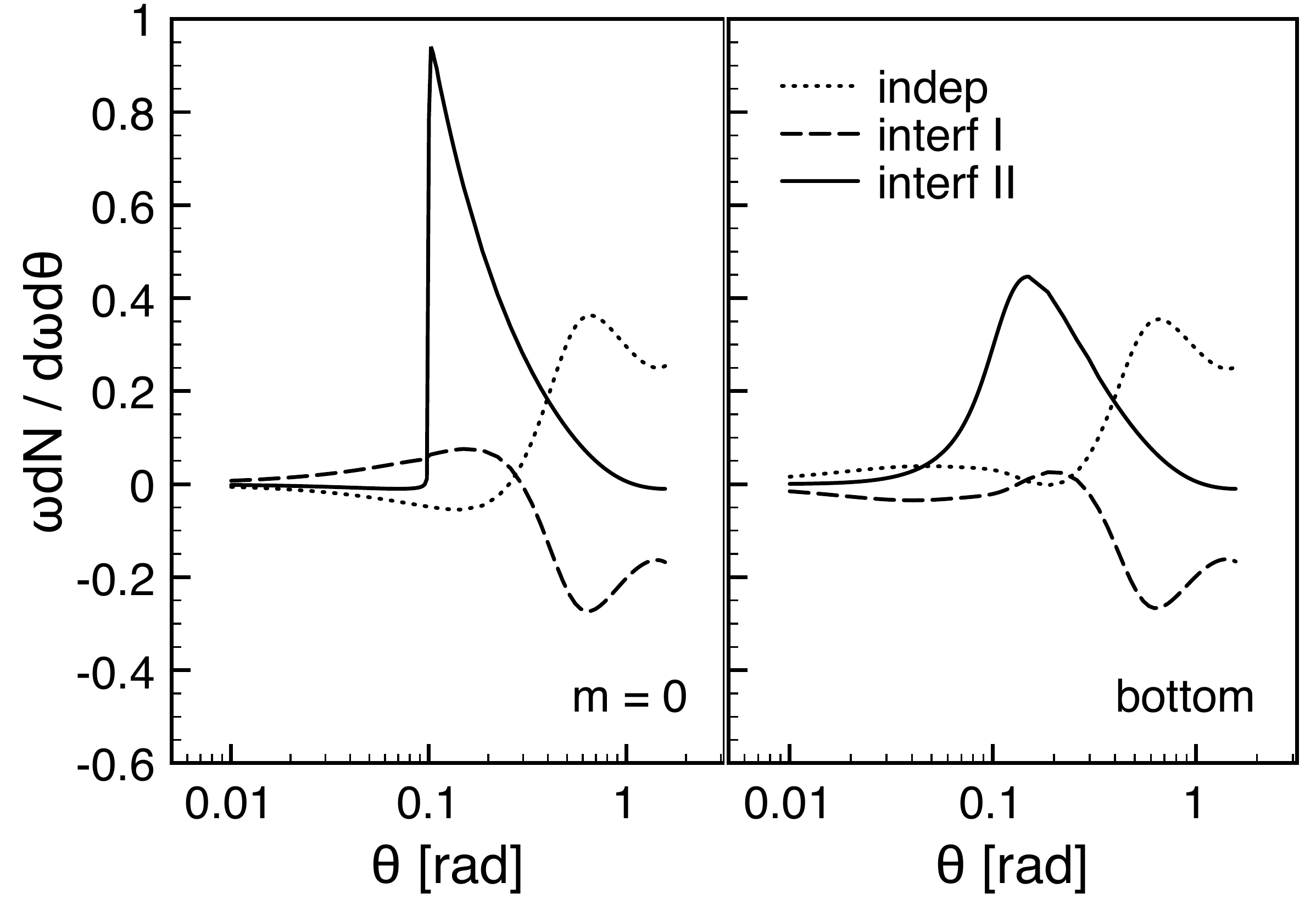}
\caption{\label{fig:GIC}Cancellation between the sum of the independent spectra for $q + {\bar q}$ and the interference spectrum I, for massless (plot on the left) and bottom ($m_q=5$ GeV, plot on the right) quarks. The quark and the antiquark energies are $E_q=E_{\bar q}=100$ GeV, the gluon energy $\omega=2$ GeV, the Debye mass $m_D=0.5$ GeV, the medium length $L=4$ fm and the antenna opening angle $\theta_{q {\bar q}}=0.1$. The dotted curve corresponds to the independent spectra for $q + {\bar q}$, the dashed curve corresponds to the interference spectrum I and the solid curve corresponds to the interference spectrum II.}
\end{center}
\end{figure}

The interference spectrum ``I" reads
\begin{equation}
\label{eq:interference}
\begin{split}
{\cal I}_{q {\bar q}}^{\rm interf \, I} = & - \, 2 \,(4\pi)^2 \alpha_s^2  \, C_A \, C_F\, n_0\, \int_0^{L^+} {\rm d} x^+  \, \int \frac{{\rm d}^2\q}{(2 \, \pi)^2} \, \frac{m _D^2}{(\q^2 + m _D^2)^2} \\
& \frac{1}{x \, {\bar x}} \bigg[\frac{{\boldsymbol \kappa} \cdot {\bar {\boldsymbol \kappa}}}{(p \cdot k) \, ({\bar p} \cdot k)} \Big(\cos\big(\Omega_{q {\bar q}}^0 \, x^+\big) - 1\Big) \\
& + \frac{{\boldsymbol \nu} \cdot {\bar {\boldsymbol \nu}}}{(p \cdot v) \, ({\bar p} \cdot v)} \Big(1 + \cos\big(\Omega_{q {\bar q}} \, x^+\big) - \cos\big(\Omega_q \, x^+\big) - \cos\big(\Omega_{\bar q} \, x^+\big)\Big) \\
& + \frac{{\boldsymbol \nu} \cdot {\bar {\boldsymbol \kappa}}}{(p \cdot v) \, ({\bar p} \cdot k)} \Big(\cos\big(\Omega_{\bar q} \, x^+\big) - \cos\big(\Omega_{q {\bar q}} \, x^+\big)\Big) \\
& + \frac{{\bar {\boldsymbol \nu}} \cdot {\boldsymbol \kappa}}{({\bar p} \cdot v) \, (p \cdot k)} \Big(\cos\big(\Omega_q \, x^+\big) - \cos\big(\Omega_{q {\bar q}} \, x^+\big)\Big)\bigg],
\end{split}
\end{equation}
with $\Omega_{q {\bar q}}$ $=$ $p \cdot v / p^+ - {\bar p} \cdot v / {\bar p}^+$ and $\Omega_{q {\bar q}}^0$ $=$ $p \cdot k / p^+ - {\bar p} \cdot k / {\bar p}^+$, and the interference spectrum ``II" reads
\begin{equation} 
\label{eq:brems1}
\begin{split}
{\cal I}_{q {\bar q}}^{\rm interf \, II} = & - \, 2 \,(4\pi)^2 \alpha_s^2  \, C_A \, C_F\, n_0\, \int_0^{L^+} {\rm d} x^+  \, \int \frac{{\rm d}^2\q}{(2 \, \pi)^2} \, \frac{m _D^2}{(\q^2 + m _D^2)^2} \\
& \frac{{\boldsymbol \kappa} \cdot {\bar {\boldsymbol \kappa}}}{x \, {\bar x} \, (p \cdot k) \, ({\bar p} \cdot k)} \Big(\cos\big(\Omega_{q {\bar q}} \, x^+\big) - \cos\big(\Omega_{q {\bar q}}^0 \, x^+\big)\Big).
\end{split}
\end{equation}
The divergent structure of Eq.~(\ref{eq:brems1}) is apparent from the fraction in the second line. Furthermore, arranged in this way, the novel contributions exhibit a cancellation between  ${\cal I}_{q {\bar q}}^{\rm indep}$ and ${\cal I}_{q {\bar q}}^{\rm interf \, I}$ in the soft limit. This cancellation, first observed in \cite{MehtarTani:2010ma}, still holds numerically in the massive case, see Fig. \ref{fig:GIC}.

In the following, we will be concerned with comparing the coherent spectrum off one of the constituents, say the quark, defined as
\beq
\label{eq:coherentspectrum}
\Icoh = \Iindep + \left( \IinterfOne + \IinterfTwo\right)/2 \,,
\eeq
with the purely independent component, $\Iindep$. In vacuum, it is well knowm \cite{Basics of Perturbative QCD} that the coherent spectrum off the quark (or the antiquark) differs from the independent spectrum only by the angular ordering condition.

For the medium-induced spectrum, on the other hand, since only the $\IinterfTwo$ component contains an infrared divergence, the main contribution to Eq.~(\ref{eq:master2}) is readily found for soft gluon emissions --- see \ref{sec:appb} for details.
In the ultrarelativistic limit, after averaging over the azimuthal angle with respect to the quark direction, we obtain
\begin{equation} 
\label{eq:ClosedForm}
\omega \frac{{\rm d} N_q  }{{\rm d} \omega \, {\rm d} \theta} = \frac{2\,\alpha_s  C_F}{\pi} \frac{\sin{\theta} }{1 - \eta \cos{\theta}} \, \frac{1+\eta^2}{1+\cos\theta} \, H (\theta_\qqb,\theta; \theta_0) \, \Delta_{\rm med} (\theta_{q {\bar q}}, \theta_0, L^+) \,,
\end{equation}
where $\eta = \sqrt{1-\theta_0^2}$. Note, that the collinear divergency in the second fraction on the right-hand-side is explicitly regulated by the mass. In Eq.~(\ref{eq:ClosedForm}), the expression for $H (\theta_\qqb, \theta;\theta_0)$ reads
\begin{equation} 
\label{eq:Hfactor}
H (\theta_\qqb, \theta;\theta_0) = \frac{1}{2}\left(1+ \frac{\eta \cos{\theta_{q {\bar q}}} - \cos{\theta}}{\sqrt{\big(1-\eta \cos\theta_\qqb \cos\theta \big)^2 - \eta^2 \sin^2\theta_\qqb \sin^2\theta}} \right)\,,
\end{equation}
which reduces to the Heaviside step function in the massless case, $H(\theta_\qqb,\theta;\theta_0=0) = \Theta \big(\cos{\theta_{q {\bar q}}} - \cos{\theta} \big)$. This generalized Heaviside function comes with the reverse ordering condition than found for vacuum radiation, thus mainly allowing radiation to be induced at angles larger than the opening angle of the pair. This spectrum is therefore a generalization of the property of antiangular ordering, found for the first time in \cite{MehtarTani:2010ma}, for massive quarks here.

The medium decoherence parameter $\Delta_{\rm med}$, appearing in Eq.~(\ref{eq:ClosedForm}), reads
\begin{equation} 
\label{eq:mediumdecoh}
\begin{split}
\Delta_{\rm med} (\theta_{q {\bar q}}, L^+) & = \frac{{\hat q}}{2 \, m_D^2} \int_0^{L^+} {\rm d} x^+ \, \bigg[1 - \frac{|{\r}| \, m_D \, x^+}{L^+} K_1 \bigg(\frac{|{\r}| \, m_D \, x^+}{L^+}\bigg)\bigg] \\
& \approx \frac{1}{12} \, {\hat q} \, L^+ \, |{\r}|^2 \bigg(\log{\frac{1}{|{\r}| \, m_D}} + {\rm const.}\bigg),
\end{split}
\end{equation}
where $K_1(x)$ is the modified Bessel function of the second kind and $|{\r}|=|\delta {\boldsymbol n}| \, L^+$ is the transverse separation between the emitters of the antenna when they leave the medium. The transverse angular separation of the pair $|\delta {\boldsymbol n}| \sim  \theta_\qqb$ is given explicitly in Eq. (\ref{eq:eqn}), and contains only a weak dependence on the quark mass. Finally, we have defined a medium transport coefficient $\hat q= 2 \alpha_s C_A n_0 m_D^2$, which differs slightly from the standard definition\footnote{The standard definition of $\hat q$ refers to a slightly different scheme, namely the multiple soft scattering approximation.} in \cite{Salgado:2003gb}.

Note that the approximation in the second line of Eq. (\ref{eq:mediumdecoh}) is strictly valid as long as $|{\r}| \, m_D$ $\ll$ $1$. Under the same condition, one can further simplify Eq. (\ref{eq:mediumdecoh}) by observing that $\Delta_{\rm med} (\theta_{q {\bar q}}, L^+)$ exhibits only a mild logarithmic dependence on $|{\r}| \, m_D$. Then one simply has
\begin{equation} \label{eq:leadingmediumdecoh}
\Delta_{\rm med} (\theta_{q {\bar q}}, L^+) \propto {\hat q} \, L^+ \, |{\r}|^2 \propto {\hat q} \, L^3 \, \theta_{q {\bar q}}^2.
\end{equation}
Keeping in mind the angular ordered vacuum contribution, it becomes evident that $\Delta_\text{med}$ controls the onset of decoherence or, in other words, the breakdown of angular ordering for soft gluons. For dense media, one has to go beyond the single scattering approach and resum multiple interactions \cite{MehtarTani:2011tz, MehtarTani:2011jw, CasalderreySolana:2011rz}. It is easy to check that $\Delta_\text{med}$ in Eq.~(\ref{eq:mediumdecoh}) is, in fact, the leading order contribution to the full result. In the opaque limit, $\Delta_\text{med} \to 1$ thus losing all sensitivity on the medium characteristics.

%Above, we have relied on the fact that $\Omega^0_\qqb \to 0$ in the soft limit. This approximation breaks down for $\omega \gg 0$, when one has to keep the factor $\cos\big(\Omega_{q {\bar q}}^0 \, x^+\big)$ in Eq.~(\ref{eq:brems1}). This gives rise to a characteristic cut-off scale of the soft spectrum which, assuming that the bulk of medium-induced gluon production is peaked at the opening angle, scales as
%\begin{equation} 
%\label{eq:cutoffscale}
%\omega_{\rm coh} \sim (\theta_{q {\bar q}}^2 \, L)^{-1} \,.
%\end{equation}
%Note that this characteristic scale does not depend on the mass. In fact, in \cite{MehtarTani:2011jw, CasalderreySolana:2011rz} it was demonstrated that $\omega_\text{coh}$ marks the cut-off above which all medium-induced interferences are suppressed. In conclusion, this signifies that the full medium-induced coherent spectrum eventually reduces to the independent one for large opening angles of the pair and large medium lengths, and most rapidly in the hard sector. 
Above, we have relied on the fact that $\Omega^0_\qqb \to 0$ in the soft limit. This approximation breaks down for $\omega \gg 0$, when one also have to keep the factor $\cos (\Omega_\qqb x^+ )$ in Eq.~(\ref{eq:brems1}). Moreover, for hard gluons one has to take into account all the components of the spectrum on equal footing. This gives rise to a characteristic cut-off scale that governs the transition between the independent and interference-dominated parts of the spectrum. This cut-off scale is independent of the mass. For further discussions, see \cite{CasalderreySolana:2011rz}.

\section{Numerical results}

In order to analyze the new interference contributions, compare them with the already known independent ones and explore phenomenological consequen\-ces, we turn to the numerical evaluation of the results. In this work we do not pursue a full exploration of the parameter space but aim at an understanding of some differences from the known results \cite{Armesto:2003jh,Djordjevic:2003zk,Gyulassy:2000er,Wiedemann:2000za} caused by the interference terms.

Due to symmetry reason, we need only to consider the emissions off one of the antenna constituents, say the quark. Its propagation establishes a preferred direction such that all azimuthal averages are performed with respect to it and, additionally, the vector of momentum exchange with the medium is perpendicular to it \footnote{These approximations are valid in the high-energy limit, where all angles are assumed to be small.}.
Furthermore, the medium density is normalized by $n_0 \, L^+=1$ and the strong coupling constant is fixed to be $\alpha_s=1/3$. If not specified explicitly, all calculations are made for quark and antiquark energies of $E_q=E_{\bar q}=100$ GeV. 
The heavy quark masses are chosen to be $m_c=1.5$ GeV for the charm quark and $m_b=5$ GeV for the bottom one.

The choice of the remaining medium parameters reflects two extreme scenarios: the ``moderate medium" interaction with $m_D= 0.5$ GeV and $L = 4$ fm and the ``dense medium" interaction with $m_D = 2$ GeV and $L = 10$ fm. In a realistic situation both of these quantities will depend strongly on local medium properties and its global evolution.

We first study the antenna angular spectrum and the decoherence features, then we analyze the transition between the antenna spectrum and the independent spectrum, to finally examine the average energy loss and the broadening --- in all cases, both the difference between the antenna and the independent case and the mass effect on them are discussed.

\subsection{The coherent spectrum and decoherence}

\begin{figure}[h]
\begin{center}
\includegraphics[width=0.9\textwidth]{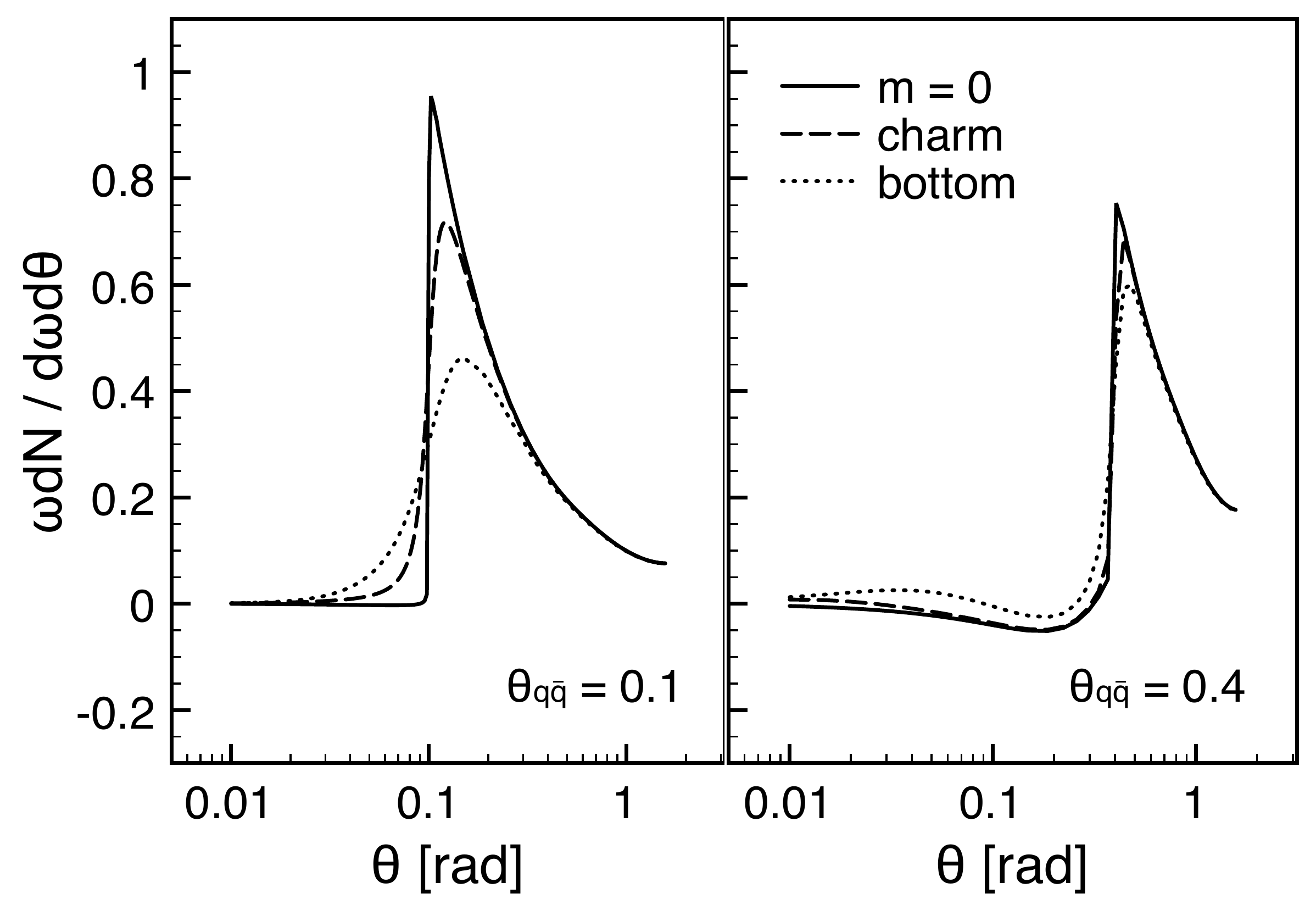}
\caption{\label{fig:Sum}The angular distribution of the medium-induced gluon radiation spectrum off a $q {\bar q}$ antenna at first order in opacity. The parameters are chosen to be the same as in Fig~\ref{fig:GIC}.
The solid curve corresponds to the massless antenna, the dashed curve corresponds to the charm antenna and the dotted curve corresponds to the bottom antenna.}
\end{center}
\end{figure}

The angular distribution of the medium-induced gluon radiation spectrum off a $q {\bar q}$ antenna at the first order in opacity is shown in Figs. \ref{fig:GIC} and \ref{fig:Sum}, where the massless antenna (solid curve) exhibits antiangular ordering for the gluon energy $\omega=2$ GeV. As already mentioned, this emerging feature is the result of cancellations between various components of the spectrum, which also hold for the massive spectrum, see Fig.~\ref{fig:GIC} and the discussion above.

For the charm and the bottom antennas, on the other hand, such strict antiangular ordering is modified by the non-zero quark (antiquark) mass for small opening angles of the pair, see right panel of Fig.~\ref{fig:GIC} and left panel of Fig.~\ref{fig:Sum}. This is due to the screening of the collinear singularity and the appearance of the generalized Heaviside function in Eq.~(\ref{eq:Hfactor}). 
The heavier the quark, the stronger the suppression of the radiation spectrum. Since the dead-cone suppression mostly affects small-angle emissions, this effect weakens significantly with an increasing opening angle of the pair, see the right panel of Fig.~\ref{fig:Sum}. In this situation, medium-induced large angle emission does not display a strong mass ordering as expected from the dead-cone suppression.

\begin{figure}[t]
\begin{center}
\includegraphics[width=0.9\textwidth]{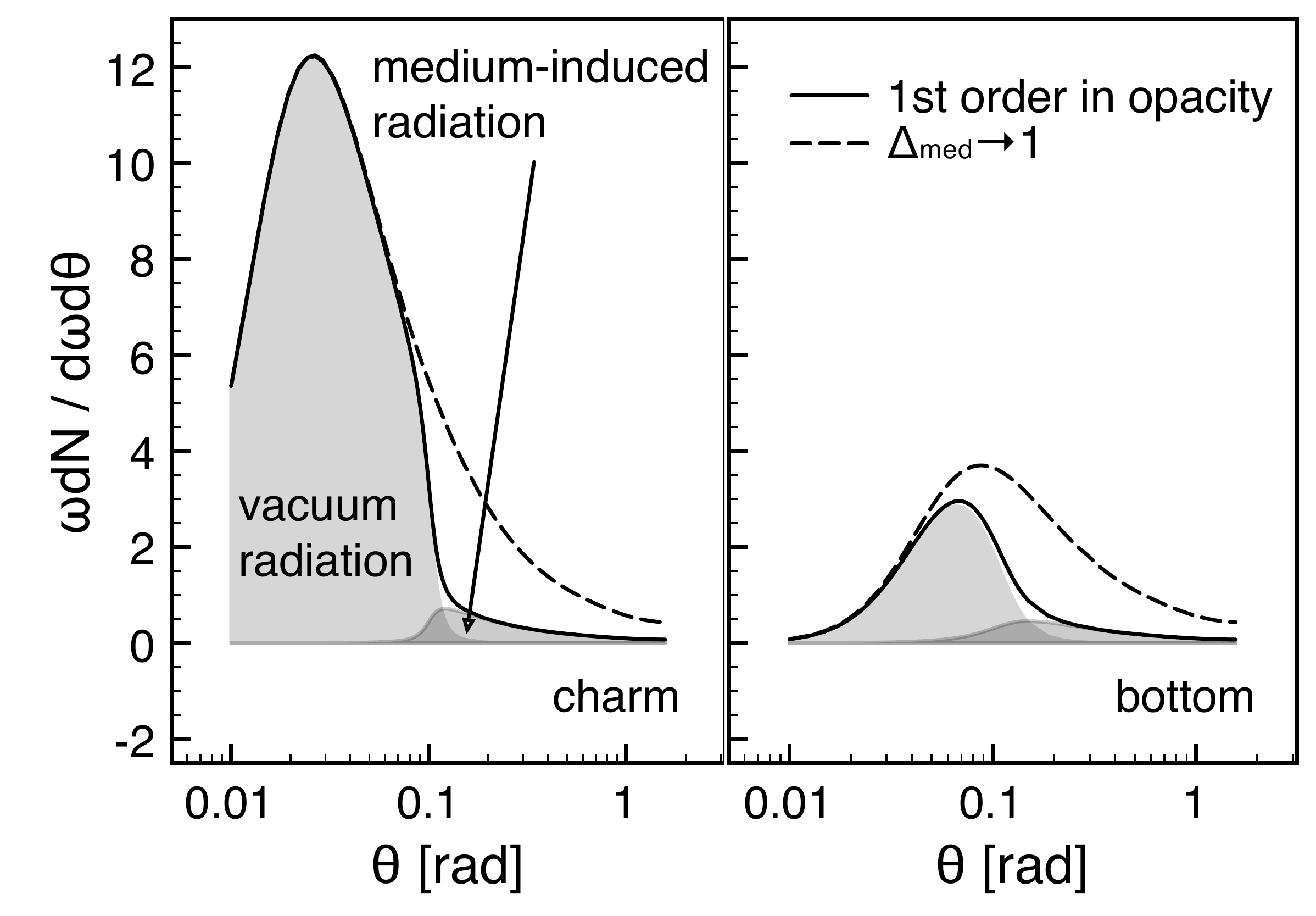}
\caption{\label{fig:Decoh}
The decoherence limit of the angular distribution of the gluon radiation spectrum off charm (plot on the left) and bottom (plot on the right) antennas in the presence of a medium. The parameters are chosen to be the same as in Fig.~\ref{fig:GIC}.
The solid curve corresponds to the spectrum computed up to first order in opacity, and the dashed curve corresponds to the antenna spectrum in the presence of a completely opaque medium, i.e., $\Delta_{\rm med}\rightarrow 1$ in Eq. (\ref{eq:ClosedForm}).}
\end{center}
\end{figure}

Both the massless antenna \cite{MehtarTani:2011tz} and the massive one exhibit complete decoherence in the opaque medium limit \cite{MehtarTani:2011tz,MehtarTani:2011jw,CasalderreySolana:2011rz}, i.e., $\Delta_{\rm med}\rightarrow 1$ in Eq. (\ref{eq:ClosedForm}), see Fig. \ref{fig:Decoh}, where we also have included the corresponding coherent vacuum spectra for charm and bottom quarks. This illustrates that the quark and the antiquark system decohere and behave like two independent emitters which radiate as if propagating in vacuum. The mass effect smears out the angular separation between vacuum and medium-induced radiation, which holds strictly in the massless case. Due to the small-angle nature of the dead-cone suppression, the intensity of the vacuum radiation is suppressed almost by a factor of 3 between the charm and bottom quark spectra while the medium-induced spectrum is almost unaffected, see also Fig.~\ref{fig:Sum}.

\begin{figure}[h]
\begin{center}
\includegraphics[width=0.9\textwidth]{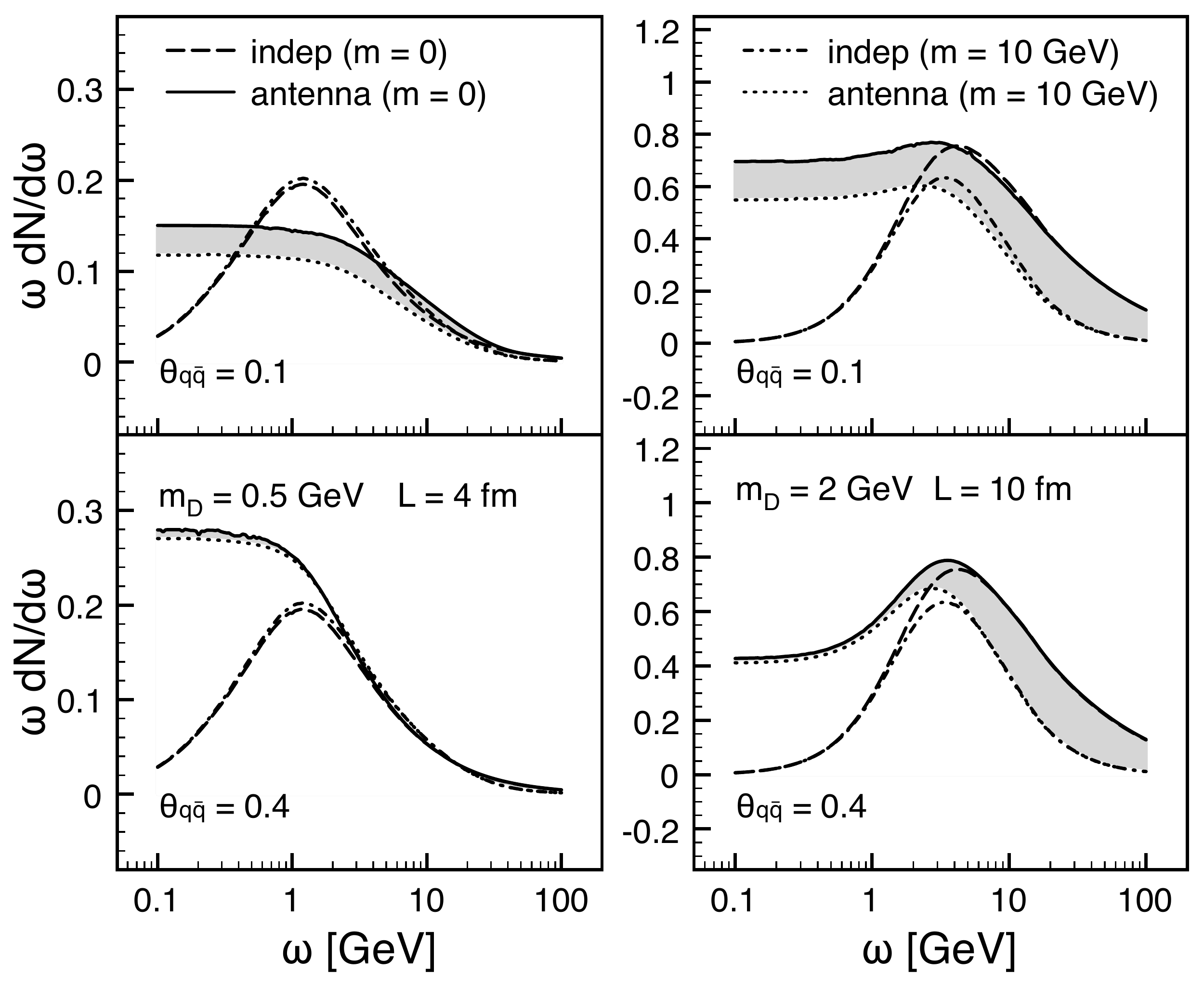}
\caption{
\label{fig:IOT}
The medium-induced gluon energy spectrum. 
We present calculations for the ``moderate medium" scenario in the left column and the ``dense medium" scenario in the right, where the opening angle is $\theta_\qqb = 0.1$ in the upper row and $\theta_\qqb=0.4$ in the lower row.
The solid curve corresponds to the massless antenna, the dotted curve corresponds to the massive antenna, the dashed curve corresponds to the massless independent spectrum and the dash-dotted curve corresponds to the massive independent one. The use of a mass of 10 GeV is for the purpose of illustration.}
\end{center}
\end{figure}

The medium-induced gluon energy spectrum is calculated via
\begin{equation}
\omega \frac{{\rm d} N}{{\rm d} \omega} = \int_0^{\pi / 2} {\rm d} \theta \, \omega \frac{{\rm d} N}{{\rm d} \omega \, {\rm d} \theta} \,,
\end{equation}
where the upper limit for the integration over $\theta$ is chosen so that $|{\k}|_{{\rm max}}=\omega$. In Fig.~\ref{fig:IOT} we plot the antenna and independent spectra off a quark with mass of 0 and 10 GeV (for the purpose of illustration), respectively. In the left column of Fig.~\ref{fig:IOT} we show calculations for the ``moderate medium", while the right column contains the same curves in the case of an ``dense medium". Let us presently point out some of the general features of these distributions, for a full discussion we refer to \cite{Armesto:2003jh,Salgado:2003gb}. First of all, we notice the different behaviors in the soft sector of the two types of spectra reflecting the infrared properties of $\Iindep$ and $\Icoh$, see Sections~\ref{sec:independent} and \ref{sec:interferences}. Secondly, we note that the coherent spectrum matches the independent one for $\omega > \omega_\text{coh}$, as expected. In between these extremes, the cancellations that dominate for dilute media, cf. upper left panel of Fig.~\ref{fig:IOT}, turn to an enhancement of the coherent spectrum, cf. lower left panel of Fig.~\ref{fig:IOT}. These general features of the medium-induced coherent spectrum holds for both (massless and massive) cases.

The effects of the quark mass are found both in the soft and in the hard sector and are highlighted by the shaded area in Fig.~\ref{fig:IOT}. Starting with the former, for the antenna spectrum this is a manifestation of the dead-cone effect which is sizeable at small opening angles, see the upper row of Fig.~\ref{fig:IOT}, and vanishes with an increasing opening angle, see the lower row of Fig.~\ref{fig:IOT}. Clearly, the dead-cone effect is more pronounced in the soft gluon regime for the antenna spectrum. The independent spectrum, on the other hand, is not noticeably modified due to the compensation between the dead-cone suppression and formation time effect. In the hard sector, we are only left with independent components, as mentioned before. Here, we observe that the spectra off massive quarks are more steeply falling than in the massless case. This is a well-known effect from the constraint on the phase space of perpendicular momentum, which ultimately gives a manifestation of the dead-cone effect. As seen in the right column of Fig.~\ref{fig:IOT}, this effect is independent of the opening angle of the pair. Finally, we note that the independent spectrum off a massive quark is in fact enhanced for small medium parameters, i.e. in the ``moderate medium" scenario in the left column of Fig.~\ref{fig:IOT}, compared to the massless case. This reflects the situation when the formation time effect prevails over the dead-cone suppression, leading to a net enhancement of medium-induced radiation, as already noticed in \cite{Armesto:2003jh}.

\subsection{Average energy loss}

\begin{figure}[t!]
\begin{center}
\includegraphics[width=\textwidth]{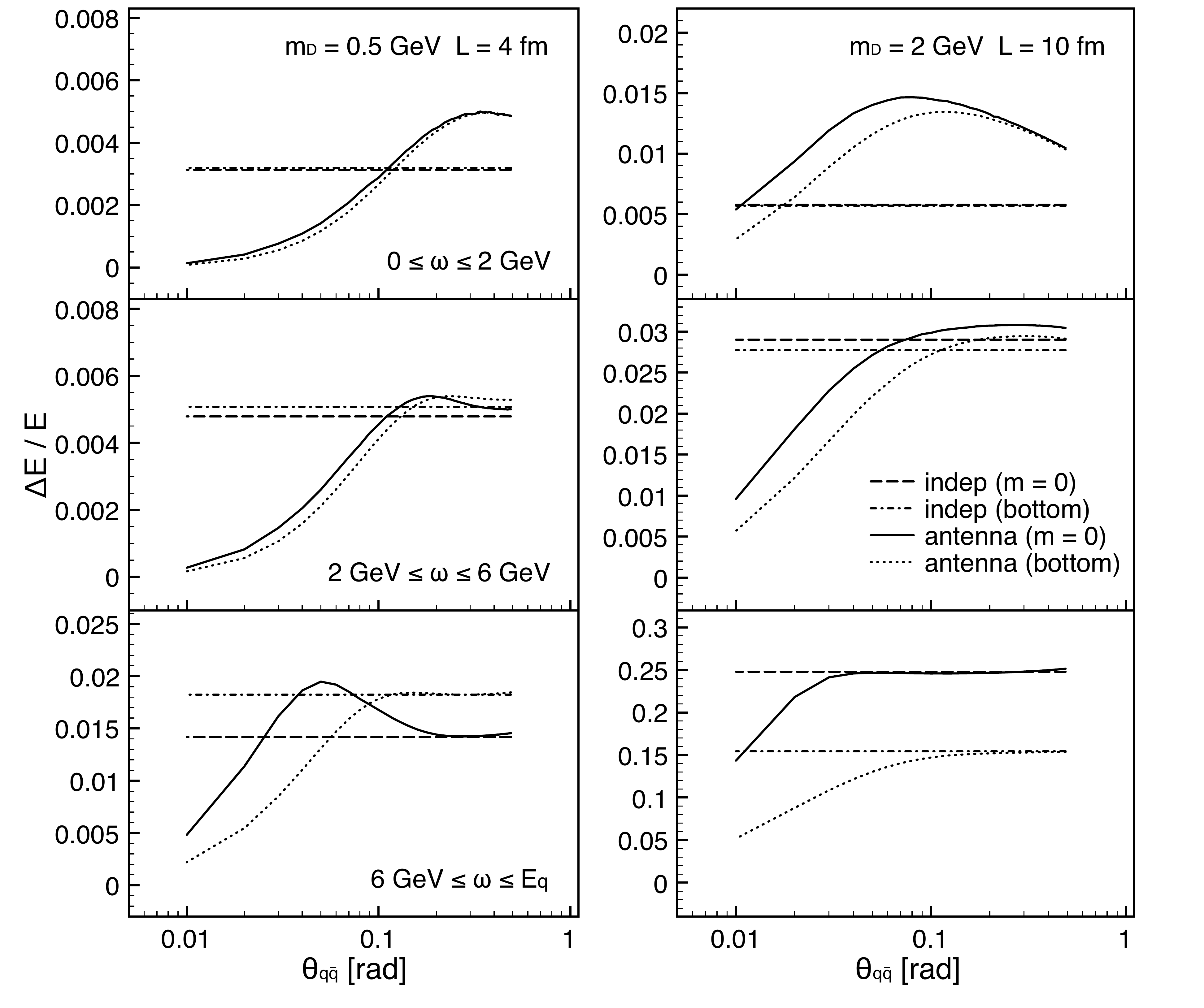}
\caption{
\label{fig:HOA}
Dependence of the medium-induced radiative relative energy loss on the antenna opening angle. The parameters are: Debye mass $m_D=0.5 \, (2)$ GeV and medium length $L=4 \, (10)$ fm for the plots on the left (right). The solid curves correspond to the massless antenna, the dotted curves to the bottom antenna, the dashed curves to the massless independent spectra and the dash-dotted curves to the bottom independent spectra. From top to bottom, the values used in Eq. (\ref{deoe}) for $\omega_{\rm min}$ are 0, 2 GeV and 6 GeV, while those for $\omega_{\rm max}$ are 2 GeV, 6 GeV and $E_q$ ($=$ $100$ GeV).}
\end{center}
\end{figure}

The amount of energy taken by the radiated gluon can be interpreted as an energy loss of the emitter. Thus, we define the radiative energy loss for the emission of gluons with energies in a certain energy interval, $\omega_{\rm min} < \omega <\omega_{\rm max}$, as
\begin{equation}
\Delta E = \int_{\omega_{\rm min}}^{\omega_{\rm max}} {\rm d} \omega \int_0^{\pi / 2} {\rm d} \theta \, \omega \frac{{\rm d} N}{{\rm d} \omega \, {\rm d} \theta}\ .
\label{deoe}
\end{equation}
The ratio $\Delta E / E$ as a function of $\theta_{q {\bar q}}$ is shown in Fig. \ref{fig:HOA} for three distinct gluon energy ranges.
In the soft and semi-soft gluon energy regions ($0\leq\omega\leq 2$ GeV and $2$ GeV $\leq \omega\leq 6$ GeV), both the massless and the massive antenna fractional energy losses grow monotonously with an increasing opening angle $\theta_{q {\bar q}}$ and the former energy loss is larger than the latter. In the hard gluon radiation sector ($6$ GeV $\leq\omega\leq E_q$), the situation is similar for large medium parameters, i.e. for the ``dense medium" scenario,
while for small medium parameters --- the ``moderate medium" scenario ---
there is a reversal of the behaviors between the massless and the bottom antennas due to formation time
effect which result in a larger energy loss for larger masses, as discussed above \cite{Armesto:2003jh}. For large medium parameters the dead-cone suppression is the main mass effect
in all gluon energy sectors. 

Both the massless and massive antenna results approach the ones from independent emitters when $\theta_{q {\bar q}}$ is large, displaying that the interference between the quark and the antiquark reduces with an increasing opening angle, as expected. Besides, in the soft gluon emission region and for large medium parameters, there is apparently more energy loss in the antenna than for independent emitters in both the massless and the massive cases. This reflects the fact that the antenna spectrum exhibits a soft divergence while the independent spectrum is infrared finite, see Fig. \ref{fig:IOT} and the discussion below. In the moderate and the hard gluon emission sectors, the antenna average energy loss increases and gradually approaches the independent average energy loss with an increasing antenna opening angle $\theta_{q {\bar q}}$, which indicates that, in general, more collimated projectiles lose less energy. Naturally, there is no medium-induced antenna radiation for $\theta_{q {\bar q}}$ $\rightarrow$ $0$ due to conservation of color. Overall, the phase space restriction for gluon radiation implied by the dead-cone effect is similar for both the antenna of large opening angles and the independent emitters.

\begin{figure}[t!]
\begin{center}
\includegraphics[width=\textwidth]{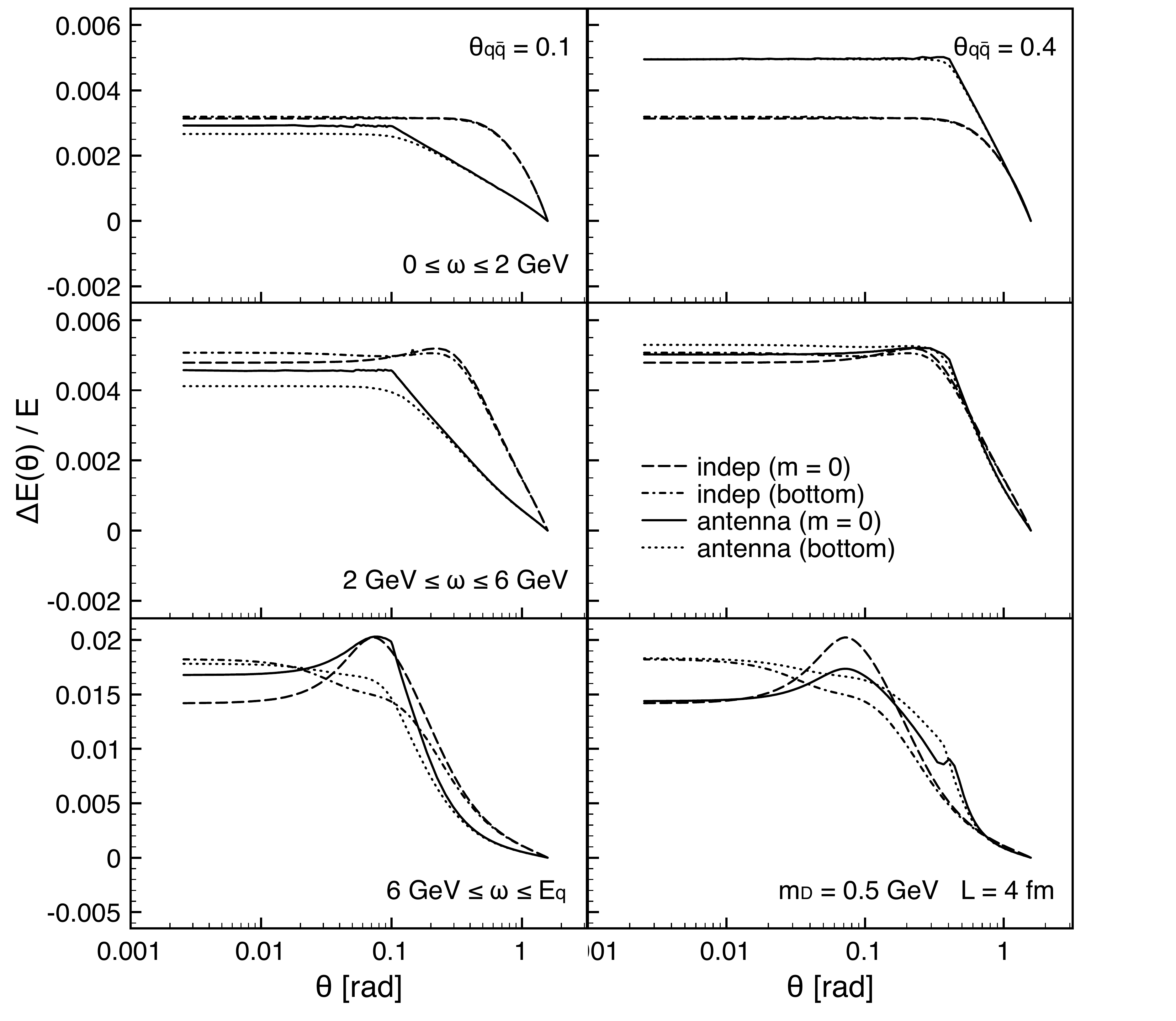}
\caption{\label{fig:HELT}Dependence of the medium-induced radiative relative energy loss outside a cone on the angle defining the cone. The parameters are: Debye mass $m_D=0.5$ GeV, medium length $L=4$ fm, and antenna opening angle $\theta_{q\bar q}=0.1\, (0.4)$ for the plots on the left (right). The solid curves correspond to the massless antenna, the dotted curves to the bottom antenna, the dashed curves to the massless independent spectra and the dash-dotted curves to the bottom independent spectra. From top to bottom, the values used in Eq. (\ref{deoep}) for $\omega_{\rm min}$ are 0, 2 GeV and 6 GeV, while those for $\omega_{\rm max}$ are 2 GeV, 6 GeV and $E_q$ ($=$ $100$ GeV).}
\end{center}
\end{figure}

\begin{figure}[h!]
\begin{center}
\includegraphics[width=\textwidth]{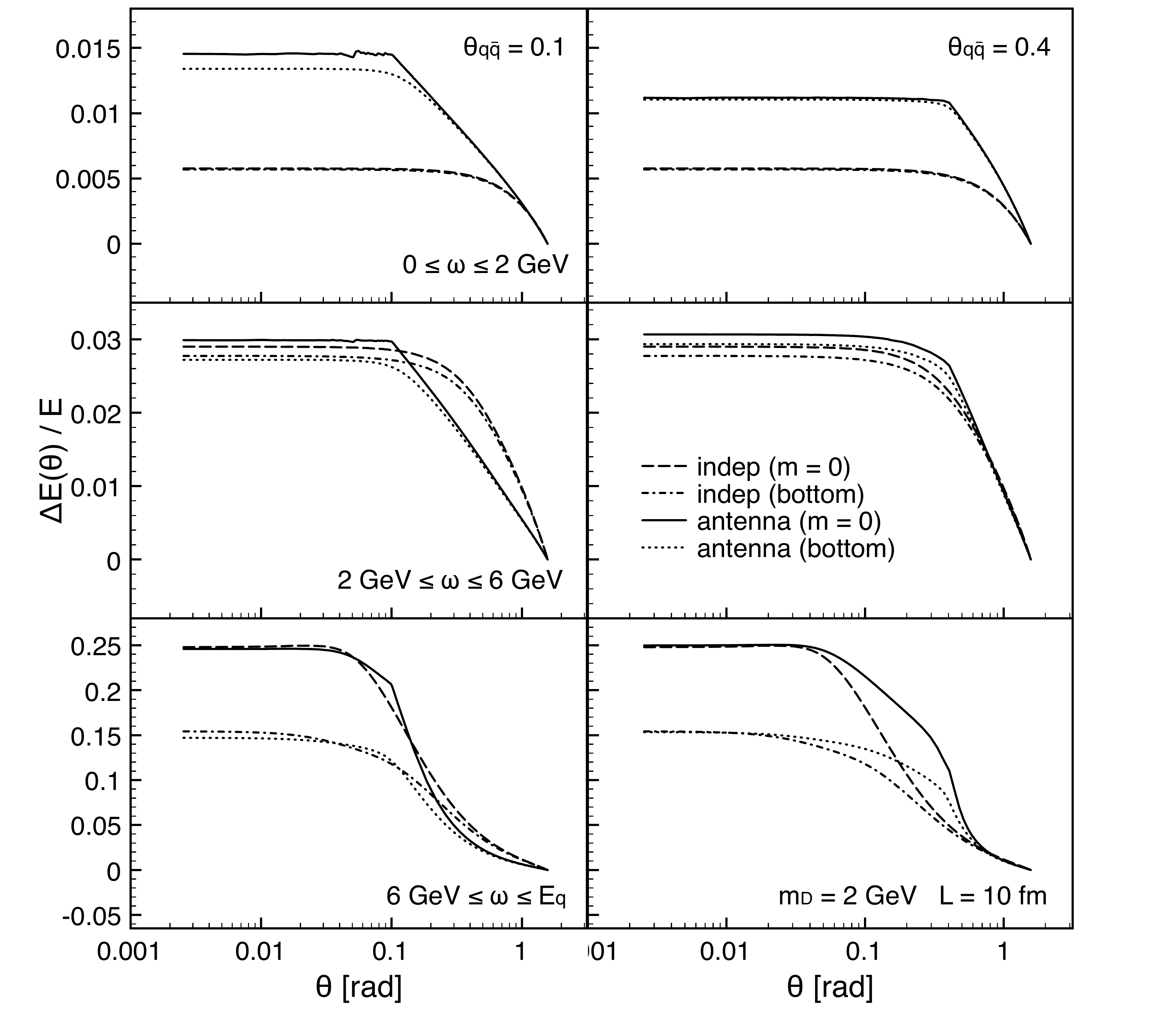}
\caption{\label{fig:HELT-10}Dependence of the medium-induced radiative relative energy loss outside a cone on the angle defining the cone. The parameters are: Debye mass $m_D=2$ GeV, medium length $L=10$ fm, and antenna opening angle $\theta_{q\bar q}=0.1\, (0.4)$ for the plots on the left (right). The solid curves correspond to the massless antenna, the dotted curves to the bottom antenna, the dashed curves to the massless independent spectra and the dash-dotted curves to the bottom independent spectra. From top to bottom, the values used in Eq. (\ref{deoep}) for $\omega_{\rm min}$ are 0, 2 GeV and 6 GeV, while those for $\omega_{\rm max}$ are 2 GeV, 6 GeV and $E_q$ ($=$ $100$ GeV).}
\end{center}
\end{figure}

In order to further investigate the relation between the average energy loss and its angular structure,
we compute the medium-induced radiative energy loss outside of a specific emission angle $\theta$:
\begin{center}
\begin{equation}
\Delta E (\theta) = \int_{\omega_{\rm min}}^{\omega_{\rm max}} {\rm d} \omega \int_{\theta}^{\pi / 2} {\rm d} \theta' \, \omega \frac{{\rm d} N}{{\rm d} \omega \, {\rm d} \theta'}\ .
\label{deoep}
\end{equation}
\end{center}
The ratio $\Delta E (\theta) / E$ as a function of $\theta$ is shown in Figs. \ref{fig:HELT} and \ref{fig:HELT-10} for the ``moderate" and ``dense" media scenarios, respectively.
Note that the behavior of $\Delta E (\theta)$ with the angle $\theta$ traces the angular behavior of the differential energy spectrum: A decrease slower than linear of $\Delta E (\theta)$ comes from an increasing energy spectrum, a decreasing linear behavior of $\Delta E (\theta)$ results from a flat energy spectrum, and a decrease stronger than linear of $\Delta E (\theta)$ traces a falling energy spectrum \footnote{A flat behavior of $\Delta E (\theta)$ indicates the null contribution, which is due to the antiangular ordering for the antenna. An increasing behavior of $\Delta E (\theta)$ indicates the existence of negative contributions at small angles due to destructive interferences as commented above --- a well-known phenomenon in the BDMPS-Z-W/GLV framework \cite{Armesto:2003jh,Salgado:2003gb} for small medium parameters.}. The first behavior points to ${\bf k}$-broadening of the radiation which reaches the upper bound at a finite angle. Let us also recall that the independent emitter case
is known to exhibit broadening \cite{Salgado:2003rv,Baier:2001qw}, and there a proportionality relation between $\langle {\bf k}^2\rangle$ and $\Delta E$ holds.

Examining Figs. \ref{fig:HELT} and \ref{fig:HELT-10}, we see that the independent emitter exhibits, in the regions of small and moderate $\omega$ (i.e., $0$ $\leq$ $\omega$ $\leq$ $2$ GeV and $2$ GeV $\leq$ $\omega$ $\leq$ $6$ GeV), ${\k}$-broadening which emerges due to the rescattering of the emitted off-shell gluon with the medium.  Since ${\cal I}_{q {\bar q}}^{\rm interf \, II}$ dominates in the soft limit $\omega\rightarrow 0$, and it only contains the on-shell gluon bremsstrahlung and the rescatterings of the quark and antiquark with the medium, there is no ${\k}$-broadening for the antenna. For $\theta_{q {\bar q}}=0.1$ in the regions of small and moderate $\omega$, $\Delta E (\theta)$ of the massless antenna is almost a constant for $\theta\leq\theta_{q {\bar q}}$ due to antiangular ordering. For $\theta>\theta_{q {\bar q}}$, the curve of the massless antenna drops monotonously and faster than linear with an increasing gluon emission angle. The suppression of the gluon radiation off the bottom antenna (dotted curve) as compared with the one off the massless antenna (solid curve) can be clearly seen at $\theta\leq\theta_{q {\bar q}}$, because most of the gluons are emitted around the opening angle. For  $\theta_{q {\bar q}}=0.4$, the antenna still keeps the interference feature --- flat behavior for $\theta<\theta_{q\bar q}$ --- in the soft gluon emission sector, but it shows some ${\k}$-broadening in the moderate gluon emission sector (more evident for large medium parameters). 

Note in Fig. \ref{fig:HELT} that the interference between emitters included in the antenna generates more gluon radiation at $\theta_{q {\bar q}}=0.4$ than at $\theta_{q {\bar q}}=0.1$ for the specific choice of the parameters, i.e., $m_D=0.5$ GeV, $L=4$ fm and $0$ $\leq$ $\omega$ $\leq$ $2$ GeV. It agrees with the medium-induced gluon energy spectrum (see the left column in Fig. \ref{fig:IOT}). In the region of large $\omega$ (i.e., $6$ GeV $\leq$ $\omega$ $\leq$ $E_q$), for both opening angles $\theta_{q {\bar q}}=0.1$ and $0.4$,  the antenna and the independent spectra exhibit similar features with respect to broadening in both the massless and the massive cases. One can see that the interference spectrum dominates  gluon radiation when the antenna opening angle is small and the emitted gluon is soft. In this case,  the antenna exhibits the new kind of broadening --- antiangular ordering; while the antenna behaves like a superposition of independent emitters when the opening angle is large and the radiated gluon is hard, and then the antenna shows the typical $\k$-broadening.

\section{Conclusions}
\begin{figure}[t]
\begin{center}
\includegraphics[width=0.8\textwidth]{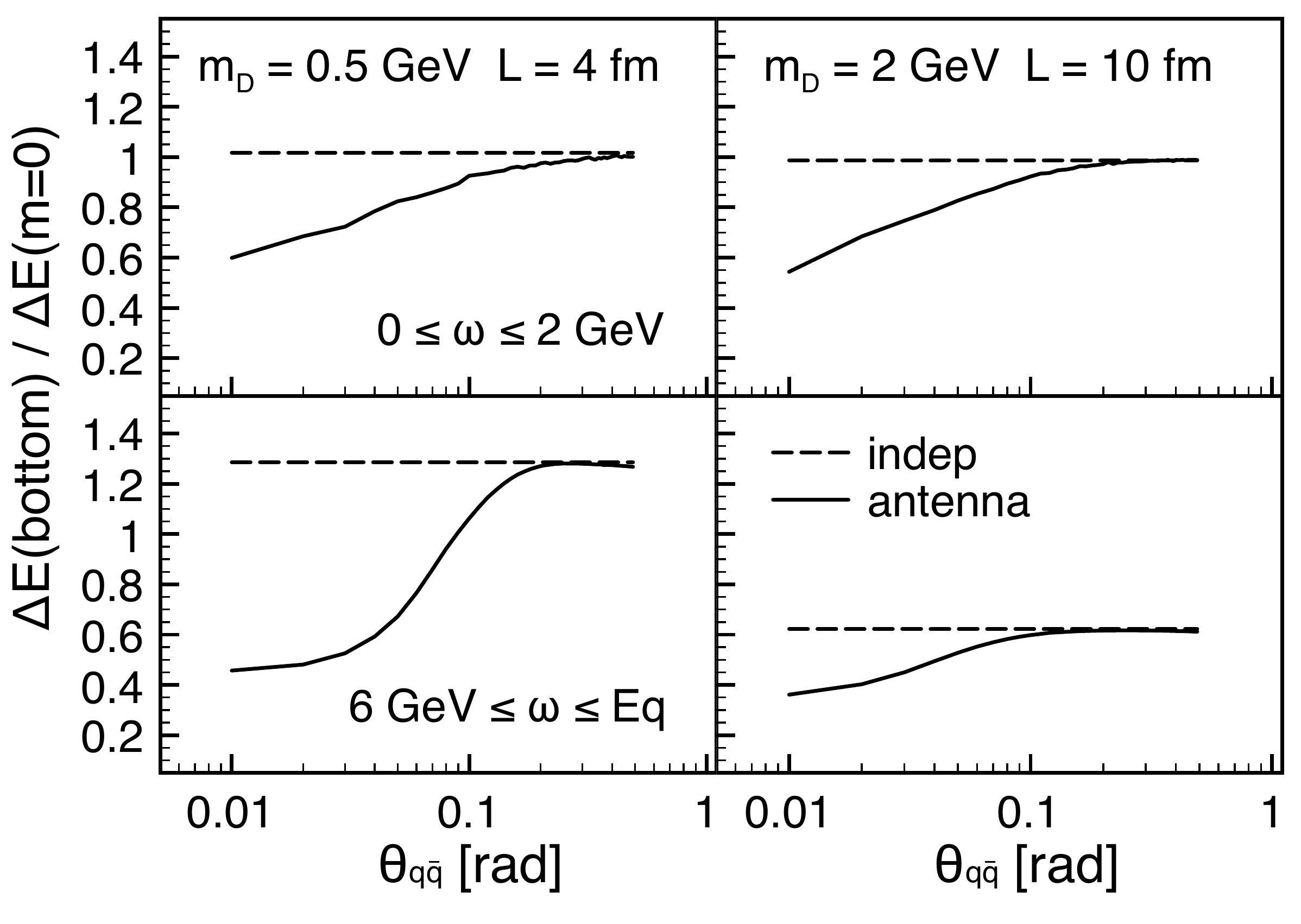}
\caption{\label{fig:summary}
Summary of mass effects for medium-induced radiation. The solid line refers to the antenna spectrum, while the dashed line to the independent one.}
\end{center}
\end{figure}

In this paper, the medium-induced gluon radiation spectrum off a $q {\bar q}$ antenna at first order in opacity is generalized to the massive case. In this calculation, performed in the high-energy limit, both the formation time effect and the dead-cone effect for massive quarks (both contained in the medium-induced gluon spectrum off individual emitting partons --- the BDMPS-Z-W/GLV formalism), and the interference between emissions off the quark and the antiquark are included. The results and techniques presented here can also be applied to tagged heavy-quark jets.

We find that the antiangular ordering obtained in the massless case \cite{MehtarTani:2010ma} is modified by the dead-cone effect. The decoherence phenomenon in \cite{MehtarTani:2011tz}  extends to the massive case. The interference between emitters included in the antenna opens the phase space for soft gluon radiation at relatively large opening angles for some specific choices of the parameters. More collimated antennas lose less energy and the size of the mass effect on the energy loss for the antenna of large opening angles is similar to the one resulting from independent emitters, see Fig. \ref{fig:summary}. The average energy loss outside of a given emission angle $\theta$ is studied, and we find that there is no typical ${\k}$-broadening in the antenna in either the massless or the massive cases for small antenna opening angles and small energies of the emitted gluon, that is the antiangular ordering regime. The antenna spectrum is found to be dominated by the contribution from independent emitters for large antenna opening angles and large energies of the emitted gluon.

The breakdown of the traditional relation between energy loss and $\k$-broadening, $\Delta E\propto \langle\k\rangle^2/L$, that we obtain in our formalism, is very suggestive of an interpretation of the recent experimental findings on reconstructed jets in nuclear collisions at the LHC \cite{Aad:2010bu, Chatrchyan:2011sx}. Indeed, the data indicate that the main observed effect is the emission of soft particles at large angles, while there is no strong modification of the fragmentation function or the dijet azimuthal asymmetry. At the same time, a large energy loss is observed in the energy imbalance between two back-to-back jets. These two features are difficult to reconcile in a traditional formalism as the relation between energy loss and broadening is a rather general one. Qualitatively, however, they admit a natural interpretation in terms of the partial decoherence which is found when more than one emitter is considered, as done here, and where vacuum-like, soft radiation, is found. Although several issues need to be clarified before  interpreting these data in terms of an underlying physical mechanism (e.g. the sample of jets studied retains some bias as demonstrated by the suppression in $R_{CP}$ measured by ATLAS), the findings presented in this paper are very encouraging for a full description of the data in terms of a medium-modified parton shower. 

\section*{Acknowledgments}

The work of NA, HM, YM-T, and CAS is supported by Ministerio de Ciencia e Innovaci\'on of Spain (grants FPA2008-01177 and FPA2009-06867-E), Xunta de Galicia (Conseller\'{\i}a de Educaci\'on and grant PGIDIT10PXIB 206017PR),  project Consolider-Ingenio CPAN CSD2007-00042, and FEDER. The work of KT is supported  by the Swedish Research Council (contract number 621-2010-3326). CAS is a Ram\'on y Cajal researcher. 

\appendix
\section{Some technical details of the calculation}
\label{sec:appa}

After performing the medium average,  the medium-induced gluon radiation spectrum at first order in opacity is given by (see \cite{Wiedemann:2000ez, Gyulassy:2000er, CasalderreySolana:2007zz})
\begin{center}
\begin{equation}
(2 \, \pi)^3 2 \, k^+ \frac{{\rm d}N}{{\rm d}^3 k_{\rm LC}} \propto \langle |{\cal M}_1|^2 \rangle + 2 \, {\rm Re} \, \langle {\cal M}_2 \, {\cal M}_0^* \rangle,
\label{master}
\end{equation}
\end{center}
where ${\cal M}_i$ is the amplitude at $i$th order in opacity expansion and $\langle \cdots \rangle$ denotes the medium average.

We use the light-cone variable $a=(a^+, a^-, {\a})$, with $a^\pm$ $=$ $(a^0 \pm a^3) / \sqrt{2}$ and the transverse coordinates ${\a}=(a^1, a^2)$ ($a^3$ $=$ $a_z$, etc.). The eikonal approximation is used in this paper, which means that the trajectory of the antenna is not altered and the only effect of the medium is a color rotation. The $q {\bar q}$ antenna moves along the $x^+$ direction and the large component of the quark (antiquark) momentum is therefore $p^+$ (${\bar p}^+$). For the gluon, we define, $k^+$ = $x \, p^+$ = ${\bar x} \, {\bar p}^+$. In the eikonal approximation, one has $E_q$ $\sim$ $E_{\bar q}$ $\gg$ $\omega$ $\gg$ $|{\k}|$ $\sim$ $|{\q}|$ (thus $x,{\bar x}\ll 1$). A light-like vector $n=(0, 1, {\bf 0})$ is defined, which specifies the light-cone gauge $n \cdot A=A^+=0$ and accordingly the medium is boosted along the $x^-$ direction i.e. the main contribution of the medium comes from the $A^-$ component. The polarization vector is decomposed as $\epsilon^* (k)=(0, {\k} \cdot {\boldsymbol \epsilon}^* / k^+, {\boldsymbol \epsilon}^*)$, which obeys the gauge \big($n \cdot \epsilon^* (k)=0$\big) and transversality \big($k \cdot \epsilon^* (k)=0$\big) conditions. The nascent $q {\bar q}$ pair originates from the splitting of a  virtual photon and then it radiates a gluon, i.e., $\gamma^*$ $\rightarrow$ $q {\bar q} + g$.

We compute the amplitude at first order in opacity (i.e., considering the contribution with one scattering center in the amplitude), but due to the unitarity, one has to take into account virtual corrections to the interaction --- the so-called contact terms --- with two scattering centers in the amplitude in the contact limit, see e.g. \cite{Gyulassy:2000er,Wiedemann:2000za}. All in all, $8\times 8$ diagrams coming from the amplitudes with one scattering plus $16 \times 2$ contact terms contribute to Eq. (\ref{master}). 

The medium is modeled as an external static potential $A_\mu(q)$. As in the BDMPS-Z-W/GLV approach, i.e. \cite{Wiedemann:2000ez, Gyulassy:2000er, Wiedemann:2000za, Baier:1996kr, Baier:1996sk, Zakharov:1996fv, Zakharov:1997uu, Gyulassy:2000fs}, we assume that the effective range of the potential corresponding to each scattering center is much smaller than the mean free path $\lambda$ of the quark (or antiquark or emitted gluon): $\lambda\gg 1 / m_D$, where $m_D$ is the Debye screening mass.  The medium average is given by \cite{Gyulassy:1993hr}
\begin{center}
\begin{equation} \label{ContactLimit}
\langle A_a^{-}(x^+, {\q}) \, A_b^{- *}(x'^+, {\q}') \rangle = \delta_{a b} \, n_0 \, m_D^2 \, \delta(x^+ - x'^+) (2 \, \pi)^2 \delta^{(2)}({\q} - {\q}') {\cal V}^2 ({\q}),
\end{equation}
\end{center}
where ${\cal V}({\q})=1/({\q}^2 + m_D^2)$ is the screened Coulomb potential.

Note that the method of computing the Feynman diagrams used in this work gives identical results to, and thus provides a cross-check of, the semiclassical method used in previous works  \cite{MehtarTani:2010ma,MehtarTani:2011tz,MehtarTani:2011jw,CasalderreySolana:2011rz}. Note also that, although in this work we develop explicitly the case of an antenna in a color singlet state, the same results hold for an antenna in an arbitrary color representation in the soft limit \cite{Basics of Perturbative QCD,MehtarTani:2010ma,MehtarTani:2011tz,MehtarTani:2011jw}.

\section{Gluon spectrum in the soft limit for small opening angles of the antenna}
\label{sec:appb}

The interference spectrum ``II''  can  be expressed as
\begin{center}
\begin{equation} \label{brems2}
\begin{split}
{\cal I}_{q {\bar q}}^{\rm interf \, II} = & \int \frac{{\rm d}^2{\q}}{(2  \pi)^2}  \int_0^{L^+} {\rm d} x^+ \, n_0  \frac{m _D^2}{({\q}^2 + m _D^2)^2} \, 2  \alpha_s^2  (4  \pi)^2  C_A  C_F \\
& \frac{{\boldsymbol \kappa} \cdot {\bar {\boldsymbol \kappa}}}{x  {\bar x}  (p \cdot k)  ({\bar p} \cdot k)} \cos\big(\Omega_{q {\bar q}}^0 \, x^+\big)  \big(1 - \cos(\delta {\n} \cdot {\q} \, x^+)\big),
\end{split}
\end{equation}
\end{center}
where the explicit expression of $\delta {\boldsymbol n}$ is
\begin{center}
\begin{equation}
\label{eq:eqn}
\delta {\n} = \frac{{\p}}{p^+} - \frac{{\bar {\p}}}{{\bar p}^+} = - \frac{\sqrt{2} \sin{\theta_{q {\bar q}}}}{1 + \sqrt{1 - \theta_0^2} \cos{\theta_{q {\bar q}}}} \, {\hat {\n}},
\end{equation}
\end{center}
with ${\hat {\bf n}}$ the direction of $\delta {\boldsymbol n}$. In the soft limit $\omega\rightarrow 0$, $\Omega_{q {\bar q}}^0$ vanishes, and the main contribution to the medium-induced gluon radiation spectrum comes from ${\cal I}_{q {\bar q}}^{\rm interf \, II}$ due to its soft divergency and to the cancellation between ${\cal I}_{q {\bar q}}^{\rm indep}$ and ${\cal I}_{q {\bar q}}^{\rm interf \, I}$ as discussed before. Then the spectrum for $\omega$ $\rightarrow$ $0$ reads
\begin{center}
\begin{equation} \label{SoftLimit}
\omega \frac{{\rm d} N}{{\rm d} \omega \, {\rm d} \theta} = \frac{8  \alpha_s  C_F  \sin{\theta} \, (1 + \sqrt{1 - \theta_0^2})}{(1 - \sqrt{1 - \theta_0^2} \cos{\theta}) (1 + \cos{\theta})} \, H (\theta_{q {\bar q}}, \theta_0, \theta) \int_{0}^{L^+} {\rm d} x^+ \, \sigma (\theta_{q {\bar q}}, \theta_0, x^+),
\end{equation}
\end{center}
where the explicit expression of $H (\theta_{q {\bar q}}, \theta_0, \theta)$ is given in Eq. (\ref{eq:Hfactor}).
The forward scattering dipole amplitude is defined as
\begin{center}
\begin{equation} \label{dipole cross section}
\sigma (\theta_{q {\bar q}}, \theta_0, x^+) = \int \frac{{\rm d}^2{\q}}{(2  \pi)^2} \, \frac{{\hat q}}{({\q}^2 + m _D^2)^2} \Big[1 - \cos{\big(\delta {\n} \cdot {\q} \, x^+ \big)}\Big],
\end{equation}
\end{center}
with ${\hat q}=2  \alpha_s  C_A  n_0  m_D^2$ the medium transport coefficient. Note that when $\theta_{q {\bar q}}\rightarrow0$, $\sigma (\theta_{q {\bar q}}, \theta_0, x^+)$ vanishes. Eq. (\ref{SoftLimit}) can be further simplified by integrating over ${\bf q}$, resulting in Eq. (\ref{eq:ClosedForm}).

\end{document}